%% file: main.tex
\documentclass[lettersize,journal]{IEEEtran}
\usepackage{amsmath,amsfonts}
\usepackage{array}
\usepackage[caption=false,font=normalsize,labelfont=sf,textfont=sf]{subfig}
\usepackage{textcomp}
\usepackage{stfloats}
\usepackage{url}
\usepackage{verbatim}
\usepackage{graphicx}
\def\BibTeX{{\rm B\kern-.05em{\sc i\kern-.025em b}\kern-.08em
    T\kern-.1667em\lower.7ex\hbox{E}\kern-.125emX}}
\usepackage{balance}

\DeclareMathOperator*{\argmax}{arg\,max}

\usepackage[ruled,vlined]{algorithm2e}
\usepackage{graphicx}
\usepackage{nccmath}
\usepackage{hhline}
\usepackage{multirow}
\usepackage{array}
\usepackage{bbm}
\usepackage{dsfont}
\usepackage{placeins}
\usepackage{float}
\restylefloat{table}

\usepackage{color}

\begin{document}

\title{Bayesian Sensor Placement for Multi-source Localization of Pathogens in Wastewater Networks}
\author{Kalvik Jakkala$^{1}$ and Srinivas Akella$^{1}$
\thanks{This document is the results of the research
   project funded in part by a legislative allocation to UNC Charlotte under the CARES Act.}
\thanks{$^{1}$The authors are with the Department of Computer Science, University of North Carolina at Charlotte, Charlotte, NC, USA. Email:{\tt\small \{kjakkala, sakella\}@charlotte.edu} Corresponding author: Kalvik Jakkala}}

\maketitle

\input{abstract}


\input{introduction}
\input{related_work}
\input{problem_statement}
\input{method}
\input{experiments}
\input{conclusion}

\section*{Acknowledgments}
We thank Saurav Agarwal and Sayantan Datta for their helpful comments. We thank Rick Tankersley for his support and the UNC Charlotte COVID wastewater team—Cynthia Gibas, Wenwu Tang, Mariya Munir, Jacelyn Rice-Boayue, Kevin Lambirth, Greg Cole, Neha Mittal, Don Chen, Jessica Schlueter, Tianyang Chen, Zachery Slocum, and other students and staff for their assistance. 

\bibliographystyle{IEEEtran}
\bibliography{references}

\appendix
\input{appendix}

\end{document}

%% file: abstract.tex
\begin{abstract}
Wastewater monitoring is an effective approach for the early detection of viral and bacterial disease outbreaks. It has recently been used to identify the presence of individuals infected with COVID-19. To monitor large communities and accurately localize buildings with infected individuals with a limited number of sensors, one must carefully choose the sampling locations in wastewater networks. We also have to account for concentration requirements on the collected wastewater samples to ensure reliable virus presence test results. We model this as a sensor placement problem. Although sensor placement for source localization arises in numerous problems, most approaches use application-specific heuristics and fail to consider multiple source scenarios. To address these limitations, we develop a novel approach that combines Bayesian networks and discrete optimization to efficiently identify informative sensor placements and accurately localize virus sources. Our approach also takes into account concentration requirements on wastewater samples during optimization. Our simulation experiments demonstrate the quality of our sensor placements and the accuracy of our source localization approach. Furthermore, we show the robustness of our approach to discrepancies between the virus outbreak model and the actual outbreak rates.
\end{abstract}

%% file: introduction.tex
\section{Introduction}

\begin{figure*}[!ht]
    \centering
	\includegraphics[width=\linewidth]{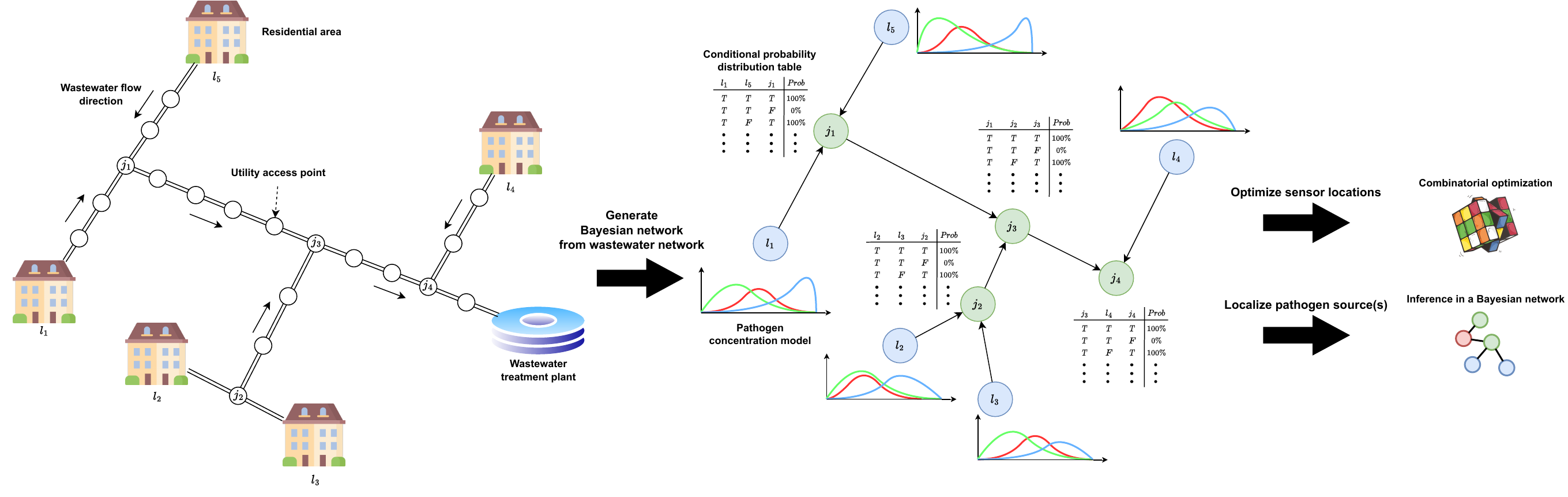}
	\caption{Illustration of our approach. We first generate a Bayesian network from the wastewater network graph. We then optimize the sensor placements using combinatorial optimization on the Bayesian network and localize the virus source(s) using inference in the Bayesian network.}
  \label{fig:overview}
\end{figure*}

There is growing evidence that the frequency of pandemics, such as the COVID-19 pandemic, is increasing~\cite{PriyadarsiniSH20, ted-ed}. It is, therefore, crucial to identify infected individuals before they show symptoms and treat them to curb the spread of such outbreaks. Wastewater-based epidemiology has proven to be an effective approach for detecting viral and bacterial outbreaks~(\cite{ChoiTDB18, SimsH20, MahonMGSM22, HughesDWWCWB22, LuoCWZZ22}), including influenza, poliovirus, respiratory syncytial virus, and Escherichia coli infections.

Recent studies have demonstrated that viruses can be detected in human waste days before an infected individual exhibits symptoms~\cite{PecciaZBGKMKMWWWWAO20}. Regular monitoring of wastewater networks for active traces of harmful pathogens can aid in identifying outbreak locations and their scale. A few case studies~(\cite{LarsenW20, KarthikeyanNMZ21, MITcovid, KisandLPPH23, WolkenSMS23, GibasLMJBRHLSYQRKNCATSM21}) have shown that monitoring wastewater networks is a reliable and non-intrusive way to detect COVID-19 outbreaks. 

Although autosamplers can automate the collection of wastewater samples, they can be costly. Additionally, the sampled wastewater must be extracted from the autosamplers and processed in a lab. As such, to keep operating costs of wastewater monitoring in check, it is crucial to minimize the number of autosamplers. It is also important to develop non-intrusive ways to quickly locate the source location(s) of pathogens that are indicators of disease outbreaks. Moreover, it is challenging to accurately detect viruses in diluted wastewater samples. Thus, we must optimize the placement locations of sensors (i.e., autosamplers) to ensure good source localization accuracy and that the sensors only collect wastewater samples of sufficiently high concentration.

However, this is a non-trivial problem that can be formulated in multiple ways. Indeed, only a few authors have recently begun to address automating sensor placement for wastewater-based epidemiology~\cite{VillezVC20, NourinejadBL21, CalleMPF21, DomokosSSTBHT22, LiHLZZY22, SpezialiMVB23}. Moreover, these approaches predominantly formulate the problem as either an area coverage problem or a probabilistic inference problem. Some authors have also incorporated graph algorithms, and \cite{SpezialiMVB23} even considered a binary Bayesian network formulation. Nonetheless, these approaches have yet to fully leverage the full potential of Bayesian networks.

In addition, given the nature of the problem, it is an insurmountable challenge to obtain any significant amount of real-world data to optimize and validate the sensor placements. This is because one would have to deploy sensors during an actual outbreak at every possible sensing location to obtain the ground truth data. Even if such data were available, finding the optimal subset of sensor placement locations is still non-trivial, as the problem is NP-hard~\cite{LeskovecKGFVG07}. 

We model the automation of wastewater-based epidemiology as a sensor placement problem. Our key contributions are summarized below:

\begin{enumerate}
    \item We present a novel Bayesian network-based formulation for the automation of wastewater-based epidemiology. Our approach can accommodate complex causal models with both discrete and continuous random variables while still being computationally feasible.
    \item Our method presents computationally efficient optimization objectives that can be solved using greedy algorithms to obtain the solution sensor placements for accurate virus source localization while also ensuring high-concentration wastewater samples are collected at the sensing locations.
    \item To validate our approach, we conduct experiments based on a real-world wastewater network with real water flow, population, and pathogen statistics, to demonstrate its accuracy and robustness.
\end{enumerate}

%% file: related_work.tex
\section{Related Work}

There has been an active response to COVID-19 in automation and robotics, involving the use of robots for disinfection, patient monitoring, deliveries, lab automation, and telemedicine (see~\cite{ShenGLMDX21, WangW21} for comprehensive surveys). However, few authors have addressed the automation of wastewater-based epidemiology.

Our problem is akin to that studied by Kempe et al.~\cite{KempeKT03}, focusing on influence maximization in social networks. In their work, the authors explored social influence networks and devised an approach to identify a subset of nodes. When influenced to take a specified action, this subset would yield the maximal number of nodes in the entire network adopting the same action. Similarly, our interest lies in garnering maximal information from the graph through the monitoring of a small number of nodes. However, our objective extends beyond mere monitoring; we aim to pinpoint information sources and model constraints on the flow of information

Berry et al.~\cite{BerryHPUW06} developed a mixed-integer programming formulation for sensor placement in water distribution networks. This method identified sensor locations to detect contaminants in water as quickly as possible. Leskovec et al.~\cite{LeskovecKGFVG07} also investigated the sensor placement problem in water distribution networks, introducing a submodular objective. Their approach efficiently optimized this objective, providing a solution with provable approximation guarantees. However, both methodologies focused on water distribution networks, which have a significantly different structure compared to wastewater networks. Moreover, neither approach addressed contaminant source localization.

Spinelli et al.~\cite{SpinelliCT19} developed an approach for source localization in physical contact networks to curb the spread of epidemics. The method took into account disease transmission times between patients to identify patient zero. However, the approach is limited to scenarios involving a single source.

Jiang et al.~\cite{JiangW06} proposed a probabilistic approach to disease outbreak detection and prediction by examining specific variables of interest that could indicate an outbreak in a given community. However, the method did not consider identifying the outbreak's source and was limited to predicting only the presence of an epidemic and its scale.

In another work, Jiang et al.~\cite{JiangNC10} developed a probabilistic spatial scan statistic that considered indicator variables to identify the presence and location of an outbreak. However, the localization approach assumed a rectilinear partitioning of the region of interest and required access to data from every sub-region. 

The source localization problem also appears frequently in other domains, including underwater localization~\cite{FangYC16}, wireless network user localization~\cite{ChengWZWLM12}, and EEG device signal localization~\cite{JatoiKMFB14}. However, most solutions to such problems involve the development of application-specific metrics and optimization objectives. We also note that our considered problem is closely related to tracking sources of pollutants in rivers~\cite{YinLZWX23}.

Peccia et al.~\cite{PecciaZBGKMKMWWWWAO20} studied the statistics of the COVID-19 virus in wastewater samples to identify general trends of community infection rates. Gibas et al.~\cite{GibasLMJBRHLSYQRKNCATSM21} examined the feasibility of sampling wastewater in a university to identify COVID-19 outbreaks and demonstrated the ability to detect asymptomatic individuals in a dormitory. However, these studies did not consider sensor placement optimization.

Recently, sensor placement approaches specifically designed for wastewater-based epidemiology have been presented in various studies~\cite{VillezVC20, NourinejadBL21, CalleMPF21, DomokosSSTBHT22, LiHLZZY22}. These approaches concentrate on optimizing sensor placement to maximize population coverage while minimizing the overlap in area coverage for each sensor, achieved through discrete integer programs.

Nourinejad et al.~\cite{NourinejadBL21} and Calle et al.~\cite{CalleMPF21} were among the first to develop sensor placement approaches for wastewater-based epidemiology. They presented probabilistic approaches to iteratively locate virus hotspots. Additionally, they \emph{separately} utilized the graph structure of wastewater networks to optimize sensor placement locations through discrete optimization.

More recently, Speziali et al.~\cite{SpezialiMVB23} formulated the wastewater sensor placement problem as an optimization task in binary Bayesian networks. However, they formulated their sensor placement approach to maximize mutual information, a computationally expensive task. Indeed, Speziali et al. resorted to quantum computing to address the computational challenges in solving their sensor placement problem.

Despite these advances, researchers in wastewater-based epidemiology have yet to fully harness Bayesian approaches, especially Bayesian networks~\cite{barber12}, capable of leveraging graph structures to model causal relationships in the problem. By employing efficient message-passing-based inference methods  for Bayesian networks, we can reduce computation costs and develop superior source localization methods.

Furthermore, our approach is not confined to simple binary networks; instead, it accommodates complex models involving multiple random variables, including continuous variables. This formulation can be solved using easily accessible traditional computers. Additionally, we present a novel Bayesian network formulation which accommodates wastewater concentration threshold requirements effectively.

%% file: problem_statement.tex
\section{ Sensor Placement for Source Localization in Wastewater Networks: Problem Statement}
We are given a wastewater network modeled as a graph $G=(V, E)$ with buildings and utility access points\footnote{Also referred to as maintenance holes, cleanouts, and sewer holes.} modeled as vertices $V$ and pipelines as directed edges $E$ whose direction is identified by the wastewater flow. Wastewater networks have a (reverse) directed tree structure with wastewater flowing from the leaf nodes (i.e., the buildings) to the root node (i.e., the wastewater treatment plant). Given their inherent graph structure, only buildings are associated with leaf nodes $l \in L$, and only utility access points are associated with non-leaf nodes $j \in J$ in the graph. We use upper case letters to represent a set of nodes, lower case letters to refer to the nodes in a set, and subscripts to indicate a specific node. 

We are given $k$ sensors (i.e., the wastewater autosamplers) that can be deployed to monitor the wastewater network. We must identify a set of nodes $W \subset V$ to monitor, i.e., select autosampler placements to accurately detect and localize virus outbreaks at one or more buildings in the wastewater network. Note that in the real world, the sensing nodes only collect the wastewater, which is then sent to a lab to be analyzed and determine if the samples are positive or negative for any viruses. Additionally, when wastewater from multiple sources is combined in pipelines, the wastewater is diluted, resulting in inaccurate virus detection test results. Therefore, we must also account for any sample concentration requirements on the wastewater while determining the solution autosampler placements.

We define a virus outbreak as a positive test for the virus of interest from the wastewater of any building. Deploying sensors at every building in the wastewater network would make detecting and localizing virus outbreaks a trivial task. However, as we are limited to only $k$ sensors (where $k << |L|$), the problem of optimizing their placement is NP-hard~\cite{LeskovecKGFVG07}, making it challenging to solve.

%% file: method.tex
\section{Approach}
We first describe our method for reducing the wastewater network graph. Then, we explain how we construct a binary Bayesian network from the reduced wastewater network graph. Next, we present our approach for using Bayesian networks to localize virus sources and optimizing the sensor placements for accurate source localization. Finally, we detail our approach to modeling more complex Bayesian networks with continuous random variables and optimizing sensor placements to meet wastewater concentration requirements. Figure~\ref{fig:overview} shows an illustration of our approach.

\subsection{Wastewater Network Graph Reduction}
\label{graph_reduce}

\begin{figure}[!ht]
    \centering
	\includegraphics[width=0.8\linewidth]{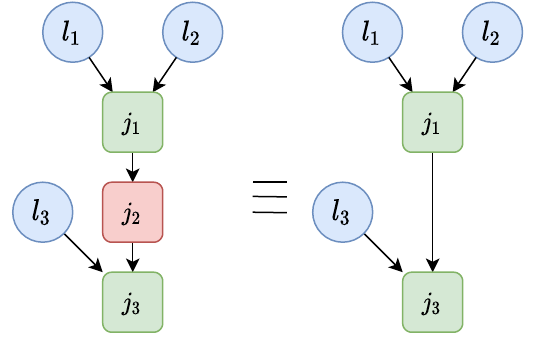}
	\caption{Node elimination scheme to reduce the graph size. Nodes with a single parent node are removed from the graph. Nodes in blue represent leaf nodes, green nodes are non-leaf nodes, and red nodes are nodes that we can remove from the graph.}
  \label{fig:parent-child}
\end{figure}

An inherent property of wastewater networks is that a virus outbreak can occur only at the graph's leaf nodes (i.e., the buildings). Furthermore, since the wastewater network graph has a (reverse) directed tree structure, we can leverage these two properties to reduce the number of nodes in the graph without losing any significant information. 

Indeed, we can exclude any non-leaf node with a single parent node, as shown in Figure~\ref{fig:parent-child}. This is because placing a sensor at either the current node $j_2$ or its parent node $j_1$ would result in sensing the same wastewater that flows into node $j_1$ from its parent nodes ($l_1$ and $l_2$). Since choosing either node ($j_1$ or $j_2$) results in the same solution, and there is no benefit from sensing at both the nodes, we discard the current node $j_2$, which is a non-leaf node with a single parent. Note that after discarding the node $j_2$, we connect the parent node $j_1$ to the child node(s) of the discarded node $j_2$. The pseudocode for the graph reduction approach is shown in Algorithm~\ref{alg:sparsify}.

Such a reduction in the graph size reduces the size of the optimization problem that needs to be solved to find the ideal sensor placements without diminishing the final solution quality. Also, it would reduce the computation cost of our source localization approach.

\subsection{Binary Bayesian Network}
\label{bgraph}

One of the main insights of this article is that we can utilize the inherent graph structure of wastewater networks to construct Bayesian graph networks. Bayesian graph networks incorporate random variables associated with each node in the graph and utilize the graph structure to model causal relationships among the random variables. We can leverage Bayesian networks to calculate conditional probabilities that can be employed to establish an optimization objective for sensor placement and also predict the most probable source(s) of a virus outbreak.

We build a Bayesian graph $B$ from the wastewater network graph $G$. In the Bayesian graph, we associate a Boolean random variable $b$ with each node. The value of $b$ indicates whether the wastewater flowing through that node contains viruses or not, where $True$ indicates the presence of viruses and $False$ indicates their absence. To construct the Bayesian graph, we must also define the distribution associated with each random variable. For the leaf nodes $l \in L$, the Boolean random variables are treated as Bernoulli distributed variables, and we parametrize the distributions to model the probability of a virus outbreak at each building.

The random variables $b$ at the non-leaf nodes $j \in J$ are computed using a deterministic OR-gate operation over the random variables of their parent nodes. Although the junction nodes $j \in J$ in real-world wastewater networks might not behave as OR-gates, this simplifying assumption allows us to efficiently compute the conditionals using variable elimination and message passing techniques~\cite{barber12}. Indeed, given our directed tree graph structure and Boolean random variables, our Bayesian graph network is similar to the Noisy-OR Bayesian network~\cite{barber12}, which is amenable to efficient Bayesian inference. In addition, our experiments show that our approach works well despite our simplifying assumption. 

\begin{figure}[!ht]
    \centering
	\includegraphics[width=0.7\linewidth]{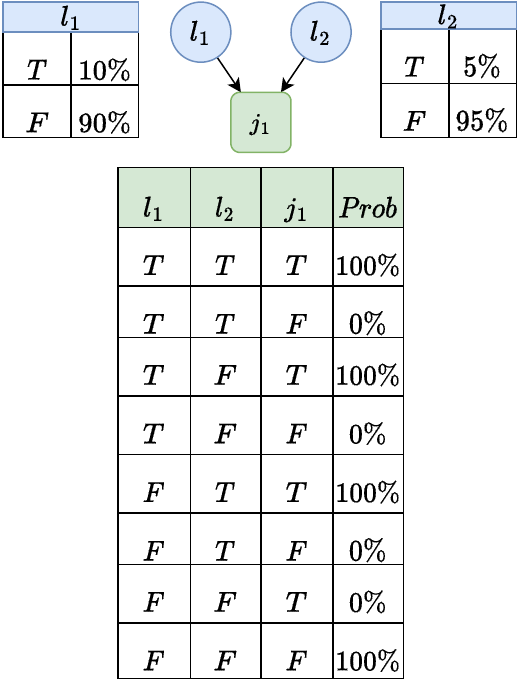}
	\caption{An example Bayesian graph $B$ for a wastewater network with two buildings. The nodes are color-coded: blue for leaf nodes $l \in L$, and green for junction nodes $j \in J$. Each CPD table's header row color and text indicate the associated node, and subscripts indicate the node index. $T$ and $F$ represent $True$ and $False$, respectively.}
  \label{fig:cpds}
\end{figure}

We assign a Conditional Probability Density (CPD) table~\cite{barber12} to each non-leaf node to store the result of the OR-gate operation. The CPD table of each node is filled by iterating over all possible instantiations of the states of the current node and its parent nodes. Each instantiation is assigned a probability of $100\%$ if the instantiation is possible with an OR-gate and set to $0\%$ otherwise, as shown in Figure~\ref{fig:cpds}.

\subsection{Source Localization}
We can now use our Bayesian graph $B$ to localize the source of virus outbreaks. Assuming we already selected a set of nodes $W \subset V$ to deploy our sensors (i.e., the wastewater autosamplers), we can use Bayesian conditioning~\cite{barber12} to predict the virus outbreak state $A_l(W_b)$ of each leaf node $l \in L$ (i.e., each building) given the current virus presence state of the sensing nodes $W_b$:

\begin{equation}
\label{preds}
A_l(W_b) =   \begin{cases}
            True &\text{if } P(b_l=True|W_b) > 50\% \\
            False &\text{otherwise}
        \end{cases} \,.
\end{equation}

Our Bayesian graph $B$ is informed about the structure of the wastewater network from the directed graph $G$, and the prior virus outbreak statistics of each leaf node from the associated Bernoulli distributions. Therefore, evaluating the most likely state $A_l(W_b)$ of each building by conditioning on the current state of the sensor nodes $W_b$ leverages all the information available to us. In addition, since our Bayesian model is a binary graph, our source localization approach is still computationally feasible.

\subsection{Sensor Placement}
\label{placement}

Our sensor placement approach leverages the Bayesian graph $B$ to formulate an optimization objective that maximizes the source localization accuracy for a given scenario $S$. This accuracy is defined as the fraction of buildings whose virus outbreak state is correctly predicted. In this context, a scenario $S$ represents a hypothetical virus outbreak indicating which of the leaf nodes (i.e., buildings) in the wastewater network are currently experiencing an outbreak.

However, we anticipate that in the real world, most scenarios involve only a small fraction of buildings experiencing a virus outbreak at any given time. As a result, even if our model predicts that all buildings are virus-free, the accuracy may still be high, as only a few buildings experiencing an outbreak are mislabeled. Therefore, optimizing and evaluating our sensor placement solutions using accuracy alone may not be the most suitable approach. As a result, we also consider precision, recall, and F1 scores as alternative evaluation metrics to determine the best optimization metric in the experiments section. In the remainder of this article, we will refer to our optimization metric as the $score$ function, implying that it can be any of the aforementioned metrics.

To ensure that our sensor placement solutions $W$ are not biased towards a single virus outbreak scenario $S$, we sample multiple scenarios, each consisting of random samples from the Bernoulli distributions associated with the leaf nodes $L$. Each scenario consists of a random sample ($True/False$) from each of the leaf nodes. Optimizing the score function on the sampled scenarios ensures that our solution sensor placements $W$ account for the probability of a virus outbreak in each building. We refer to this objective as the \emph{score objective}:
\begin{equation}
\label{opt}
\argmax_{W \subset V, |W|\leq k} \mathbb{E}_{S} \left[ score(A(W_b), S) \right] \,.
\end{equation}

Here, $A(W_b)$ represents our predicted virus outbreak state for all the buildings, and $S$ represents a virus outbreak scenario. Note that we can optimize the sensor placements by directly maximizing the score function since we have a discrete combinatorial optimization problem. Such optimization problems do not require differentiable operations. However, the above optimization problem is NP-hard~\cite{LeskovecKGFVG07}, making it challenging to find the globally optimal solution. As a result, we use the naive greedy algorithm~\cite{Minoux78} to find the solution sensor placements. The greedy algorithm selects one sensing node at a time until the cardinality constraint $k$ is met. Each new sensing node is selected by computing the increments in the optimization objective upon adding each candidate node to the current solution set and selecting the node that results in the largest increment.
\begin{equation}
\label{delta}
W \leftarrow W \cup \{ \argmax_{v \in V \backslash W} \mathcal{F}(W \cup \{v\}) - \mathcal{F}(W) \}\,.
\end{equation}

Here, $\mathcal{F}$ is the objective function (Equation~\ref{opt}). A drawback of optimizing the score objective is that it results in sensors placed only at the leaf nodes with the highest probability of an outbreak. Since we have a limited number of sensors, usually $k<<|L|$, using such a solution entails ignoring outbreaks in buildings with a lower probability of an outbreak.

To address the issue mentioned above, we have added an indicator function to the objective. This function, known as the coverage indicator function $\mathds{1}_{\text{cov}}$, returns a value of 1 if a scenario $S$ can be detected with the current sensor placements $W$, and 0 if the scenario cannot be detected. To evaluate the indicator function, we check if a path exists from the buildings experiencing a virus outbreak to the current sensor placements $W$. We refer to this objective as the Coverage objective, which can be formulated as follows:

\begin{eqnarray}
\argmax_{W \subset V, |W|\leq k} \mathbb{E}_{S} \left[ score(A(W_b), S) + \mathds{1}_{\text{cov}}(W, S)\right] \,,
\label{opt_ind}
\end{eqnarray}

\begin{equation}
\mathds{1}_{\text{cov}}(W, S) =  \begin{cases}
                    1 &\text{if scenario $S$ can be detected} \\
                       &\text{with sensors at $W$} \\
                    0 &\text{otherwise}
                    \end{cases}\,.
\label{ind_fn}
\end{equation}

By optimizing the Coverage objective, we can obtain solutions with sensing nodes capable of detecting outbreaks even at buildings with a low outbreak probability, although with a reduced source localization accuracy.

\textbf{Adding sensing nodes:} The methods discussed so far have only considered scenarios where there are no preexisting sensors deployed in the wastewater network. However, in practice, one might want to add additional sensing nodes to improve the source localization accuracy. We can achieve this by maximizing the following objective function, where $W_{curr}$ represents the set of preexisting nodes in the wastewater network, and $V \backslash W_{curr}$ represents the set of nodes in $V$ that are not in $W_{curr}$:

\begin{equation*}
\setlength{\jot}{-12pt}
\begin{split}
\argmax_{\substack{W \subset V\backslash W_{curr}, \\|W_{curr} \cup W|\leq k}} \mathbb{E}_{S} [ score&(A(\{W_{curr} \cup W\}_b), S) \\
& + \mathds{1}_{\text{cov}}(W_{curr} \cup W, S) ]
\end{split}\,.
\end{equation*}

\textbf{Removing sensing nodes:} One might also want to remove a specified number of sensing nodes from a wastewater network. This can happen if we want to monitor fewer sensors to reduce costs. We can achieve this with our approach by maximizing the following objective:

\begin{equation*}
\setlength{\jot}{-12pt}
\begin{split}
\argmax_{\substack{W \subset W_{curr}, \\|W_{curr} \backslash W|\leq k}} \mathbb{E}_{S} [ score&(A(\{W_{curr} \backslash W\}_b), S) \\
& + \mathds{1}_{\text{cov}}(W_{curr} \backslash  W, S) ]
\end{split}\,.
\end{equation*}

Note that both the problems of adding and removing sensing nodes can be solved using the greedy algorithm. However, to remove samplers, the greedy algorithm needs to be modified. In this variant, the algorithm selects the node that contributes the \emph{smallest} increment to the total objective in each iteration, unlike the sensor addition variant, which picks the node with the largest increment.

\subsection{Concentration Requirements}
\label{conc}

The methods discussed above assume that sensing nodes always report the correct state of the nodes, i.e., whether the wastewater passing through them contains viruses. However, this may not be the case in the real world. The qRT-PCR test, commonly used to analyze wastewater samples for traces of viruses, is sensitive to the concentration of the samples. Therefore, it is crucial to ensure that the concentration, measured by the number of RNA virus copies per liter of wastewater, is above a minimum threshold to obtain reliable sensing results.
\begin{figure}[th]
    \centering
	\includegraphics[width=0.7\linewidth]{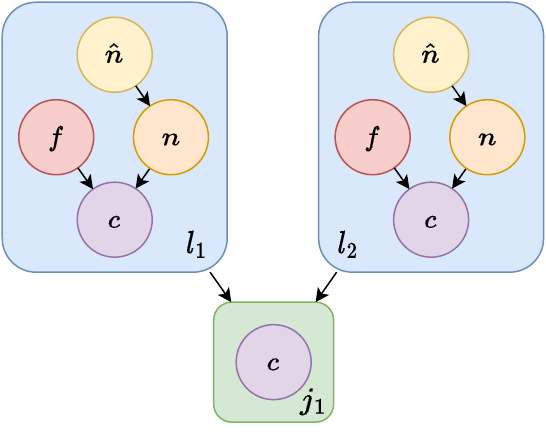}
	\caption{The concentration Bayesian graph $C$ for an example wastewater network. The nodes are color-coded as follows, red nodes: wastewater flow volume $f$, yellow nodes: number of infected individuals $\hat{n}$, orange nodes: number of total virus copies $n$ shed from each building, and purple nodes: wastewater concentration $c$. The blue and green nodes are the leaf and non-leaf nodes, respectively.}
  \label{fig:BN}
\end{figure}

However, if one were to model all variables that influence the concentration of wastewater samples, such as the volume of wastewater and the number of virus copies, in the Bayesian graph $B$, the conditional probability $P(b_l|W)$ used for sensor placement and source localization would be computationally intractable. Therefore, we construct a separate auxiliary Bayesian graph $C$, which models the variables that affect virus concentration (as shown in Figure~\ref{fig:BN}). The auxiliary graph $C$ allows us to efficiently compute the virus concentration at any node in the wastewater network, given the virus concentration at the leaf nodes $l \in L$ that are experiencing an outbreak in a scenario $S$.

The auxiliary graph $C$ is initialized with the wastewater network's graph structure similar to the binary Bayesian graph $B$. However, instead of using Boolean random variables to model each node's virus outbreak state, the auxiliary graph $C$ employs multiple random variables to represent the virus concentration at each node. Our approach considers the volume of wastewater flow $f$, the number of infected individuals $\hat{n}$, and the total number of virus copies $n$ for each building. Note that this approach can even use a more sophisticated concentration model.

We modeled the wastewater flow volume $f$ of each building using a truncated Gaussian distribution, truncated at $0$ to ensure that only positive flows are sampled. The number of infected individuals $\hat{n}$ was modeled using a Poisson distribution, while the number of total virus copies $n$ at each building was computed by sampling the number of virus copies shed by each infected individual from a uniform distribution and then taking the sum. The wastewater virus concentration $c_l$ at each leaf node $l \in L$ can be computed using the ratio of the total number of virus copies $n$ to the wastewater flow volume $f$. To determine the concentration $c_j$ at each non-leaf node $j \in J$, we use the conservation of mass equation shown below:

\begin{equation}
c_j = \frac{\sum_{i \in parent(j)}n_i}{\sum_{i \in parent(j)} {f_i}}\,.
\label{conv_mass}
\end{equation}

The concentration rate $c$ calculated at each node is then passed to an updated threshold indicator function $\mathds{1}_{\text{thresh}}$, which enforces the minimum concentration requirement in our sensor placement optimization objective. We refer to this new optimization objective as the \emph{thresholded coverage objective}.

\begin{eqnarray}
\argmax_{W \subset V, |W|\leq k} \mathbb{E}_{S} \left[ score(A(W_b), S) + \mathds{1}_{\text{thresh}}(W, S)\right]
\label{opt_ind_fn_conc}
\end{eqnarray}

\begin{equation}
\mathds{1}_{\text{thresh}}(W, S) =  \begin{cases}
                    1 &\text{if scenario $S$ can be detected} \\
                      &\text{with sensors at $W$ and every} \\
                      &\text{wastewater sample satisfies} \\
                      &\text{the concentration threshold} \\
                    0 &\text{otherwise}
                    \end{cases}
\label{ind_fn_conc}
\end{equation}

The threshold indicator function $\mathds{1}_{\text{thresh}}$~(Eq.~\ref{ind_fn_conc}) is similar to the coverage indicator function $\mathds{1}_{\text{cov}}$~(Eq.~\ref{ind_fn}), but it also considers the concentration requirements for detecting a scenario. That is, even if a path exists from the buildings experiencing a virus outbreak to the current sensor placements $W$, the threshold indicator function $\mathds{1}_{\text{thresh}}$ will return $1$ (i.e., $True$) only if the concentrations at the sensing nodes that detect the scenario meet the required concentration threshold. 

By optimizing the thresholded coverage objective, we can obtain sensor placements that not only enable accurate source localization but also ensure that the concentration of viruses in wastewater samples collected by the autosamplers is above the required threshold for reliable virus detection. It is worth noting that our approach still benefits from the computational efficiency of the Boolean OR-gate Bayesian network $B$ used to compute $P(S|W)$, while also accounting for the complex concentration requirements through the auxiliary Bayesian network $C$ and Eq.~\ref{ind_fn_conc}. Thus, our approach overcomes the limitations of a computationally intractable Bayesian graph that models all variables together.

Note that the Bernoulli distributed random variable $b_l$ in the Bayesian graph $B$, which indicates if an infection outbreak occurred at a building, is analogous to the Poisson distributed random variable $\hat{n}$ in the auxiliary Bayesian graph $C, $ which represents the number of infected individuals. Therefore, when using the auxiliary Bayesian graph $C$ to model the virus concentration, if we need to sample outbreak scenarios $S$ to optimize our sensor placements, we sample them using the Poisson distributions associated with the random variable $\hat{n}$ instead of the Bernoulli distributions associated with the random variable $b$. But since the scenarios $S$ sampled from Poisson distributions would indicate the number of infected individuals instead of the binary state of an outbreak, we apply the following \textit{min} operation on each element of $S$, i.e., the sampled $\hat{n}$ at each building to ensure that the scenario is binary: 

$$
f(\hat{n}) = \text{min}(1, \hat{n})\,.
$$

Converting each scenario $S$ into a binary vector enables us to use the binary Bayesian graph $B$ to evaluate $P(b_l|W)$ for sensor placement and source localization. However, it is important to note that the threshold indicator function $\mathds{1}_{\text{thresh}}$ is still computed using only the auxiliary Bayesian graph $C$. Additionally, when computing the wastewater concentrations, we sample the number of virus copies $n$ at a leaf node $l$ only if the corresponding building is indicated to be experiencing a virus outbreak in the scenario $S$. However, we always sample the wastewater flow volume $f$ from all buildings to account for wastewater dilution.

%% file: experiments.tex
\section{Simulation Experiments}
\label{exp}

\begin{figure}[!ht]
    \centering
	\includegraphics[width=0.8\linewidth]{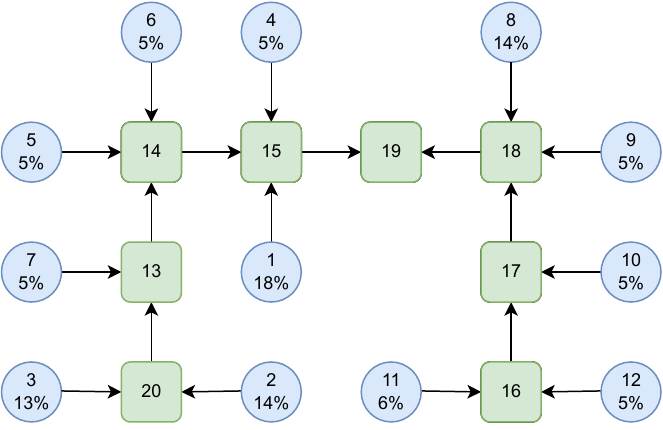}
	\caption{Subgraph of our university's wastewater network after reduction. The percentages in the leaf nodes (blue circles) indicate the fraction of the total network population at each node.}
\label{fig:sovi_blank}
\end{figure}

We tested our approach by analyzing a subgraph of our university's wastewater network consisting of residential buildings\footnote{Some aspects of the wastewater network have been obfuscated for security and confidentiality.}. Initially, we constructed the reduced graph representation of the wastewater network, denoted as $G$, by following the approach outlined in Section~\ref{graph_reduce}. The original wastewater network graph contained $35$ vertices and $34$ edges. However, we were able to reduce it to a $20$ vertex and $19$ edge graph using our graph reduction approach (shown in Figure~\ref{fig:sovi_blank}). This corresponds to a $41\%$ reduction in the number of vertices and a $44\%$ reduction in the number of edges of the graph. Out of the $20$ vertices, $12$ represented leaf nodes which corresponded to buildings.

Next, we utilized the wastewater network graph $G$ to construct two Bayesian graph networks: $B$ to model virus presence states, and $C$ to model wastewater concentrations. Our methodology for constructing $B$ and $C$ is described in Section~\ref{bgraph} and Section~\ref{conc}, respectively.

We modeled the wastewater flow volume $f$ of each building in the Bayesian graph $C$ using truncated Gaussian distributions, which were parameterized based on the buildings' historical monthly wastewater flow rates. The Poisson distribution was used to model the number of infected individuals, and the mean of this distribution was set to be proportional to the number of students assigned to the corresponding building. Lastly, we bounded the uniform distribution over the daily number of virus copies shed in the wastewater by each infected individual between $2.4\times10^6$ and $4\times10^{10}$, based on the findings of Foladori et al.~\cite{FoladoriCSMPMBR20}.

To generate hypothetical scenarios, we utilized Poisson distributions in the auxiliary graph $C$. We generated a total of $1000$ scenarios, some of which modeled simultaneous virus outbreaks in multiple buildings. These scenarios were binarized as described in Section~\ref{conc}. We used these scenarios in all of our experiments, unless specified otherwise, to maintain consistency in our benchmark results.

\subsection{Optimization Metric Benchmark}
\label{metric_benchmark}

\begin{figure}[!ht]
    \centering
	\includegraphics[width=\linewidth]{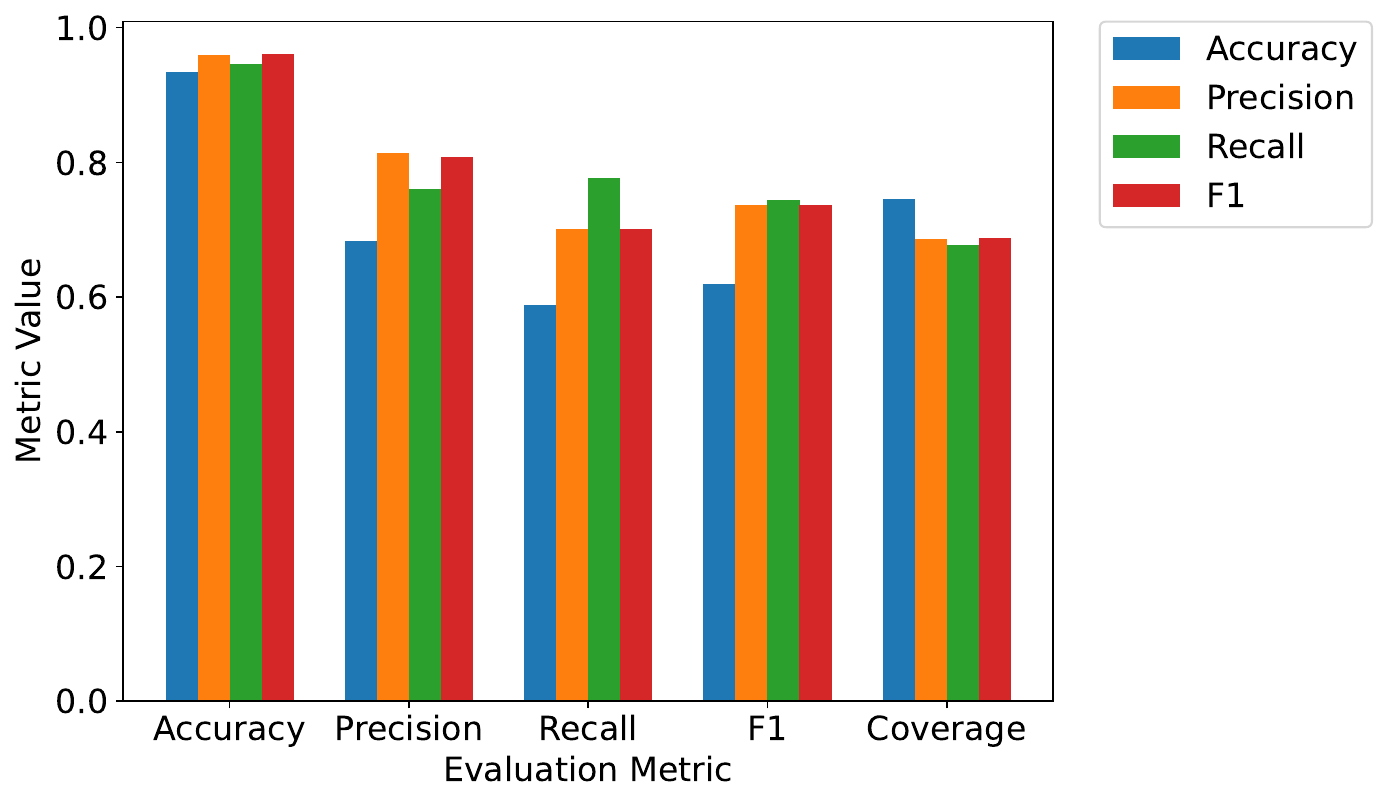}
	\caption{Optimization metric benchmark.}
\label{fig:metric_benchmark}
\end{figure}

We began by benchmarking various score functions in Equation~\ref{opt_ind_fn_conc} which is used to determine sensor placements. As discussed in Section~\ref{placement}, accuracy is not the most effective optimization metric for our approach. Therefore, we employed the greedy algorithm to optimize sensor placements using three other score functions: precision, recall, and F1 in addition to accuracy. We set the virus concentration in the threshold indicator function (Equation~\ref{ind_fn_conc}) to $4.8\times10^{5}$ and generated solutions for $6$ sensor placements. Figure~\ref{fig:metric_benchmark} displays the quality of the sensor placements obtained by optimizing with each score function. We evaluated the solution placements using all four score functions and reported them, along with the fraction of scenarios that could be covered (Equation~\ref{ind_fn_conc}) using the solution placements.

As expected, we achieved the best results for each score function when the optimization metric matched the evaluation metric. In a real-world scenario, the appropriate optimization function can be selected based on the fraction of false positive and true negative source localization predictions that can be accommodated. For the remaining experiments, we utilized only the F1 score as the optimization score function since optimizing this score provided us with good precision and recall during evaluation, even though they were not directly optimized.

\subsection{Optimizer Benchmark and Submodularity}

\begin{figure}[!ht]
    \centering
	\includegraphics[width=\linewidth]{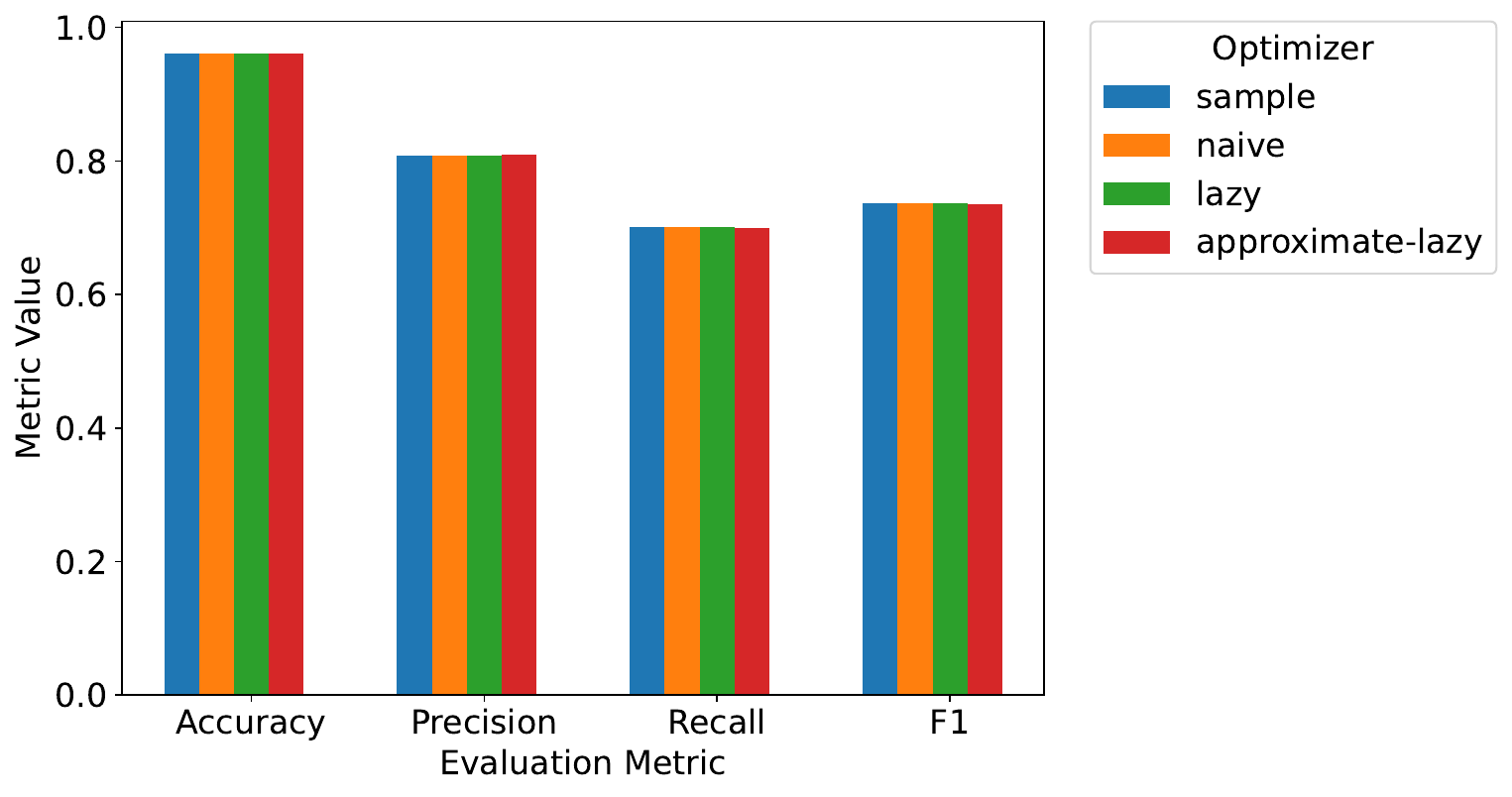}
	\caption{Optimizer benchmark.}
\label{fig:opt_benchmark}
\end{figure}

In addition to using different score functions, our approach also allows for the use of different optimizers. We benchmarked four optimizers—naive~\cite{Minoux78}, lazy~\cite{Minoux78}, approximate-lazy~\cite{WeiIB14}, and sample~\cite{MirzasoleimanBKVK15}. We generated solutions for six sensor placements using a virus concentration of $4.8\times10^{5}$ in the threshold indicator function (Equation~\ref{ind_fn_conc}), the results are shown in Figure~\ref{fig:opt_benchmark}.

The lazy, approximate-lazy, and sampls optimizers offer faster solutions than the naive optimizer without sacrificing quality, and provide a near-optimal ${(1-1/e)}$ approximation factor guarantee. This is because they assume that the optimization objective is submodular~\cite{NemhauserWF78}. We can observe that all of these approaches perform similarly to the naive optimizer, which is possible only if our optimization objective is submodular.

\subsection{Weighted Sum Benchmark}

In our previous experiments, we used an unweighted sum of the score function and the indicator function value in our optimization objective (Equation~\ref{opt_ind_fn_conc}). However, in a real-world scenario, it may be necessary to prioritize the score function or the scenario coverage (with the indicator function). As such, we introduce a weight term $\lambda$ to take a weighted sum of the score and indicator function values during optimization:

\begin{equation*}
\setlength{\jot}{-8pt}
\begin{split}
\argmax_{W \subset V, |W|\leq k} \mathbb{E}_{S} [ &(\lambda) score(A(W_b), S) \\
& + (1 - \lambda) \mathds{1}_{\text{thresh}}(W, S)]
\end{split}\,.
\end{equation*}

\begin{figure}[!ht]
    \centering
	\includegraphics[width=\linewidth]{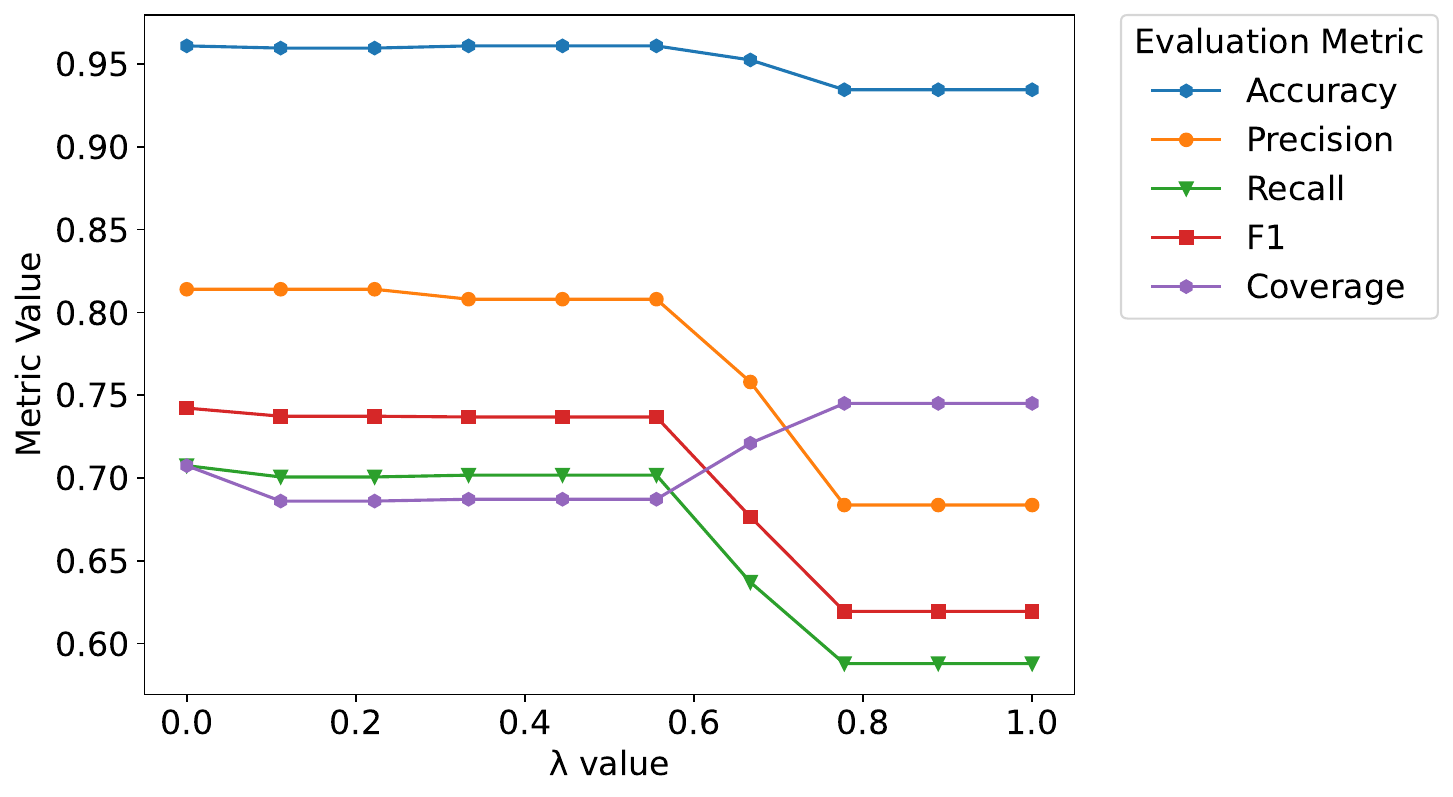}
	\caption{Weighted Sum Benchmark.}
\label{fig:lambda_benchmark}
\end{figure}

To illustrate the impact of different $\lambda$ values, we benchmarked it by generating solutions using various $\lambda$ values, and the results are presented in Figure~\ref{fig:lambda_benchmark}. We utilized the naive greedy optimizer~\cite{Minoux78}, set the virus concentration in the threshold indicator function (Equation~\ref{ind_fn_conc}) at $4.8\times10^{5}$, and produced solutions for $6$ sensor placements.

We observe from the plot that optimizing solely for coverage using the indicator function (left side) yields high score values, but the scenario coverage is somewhat reduced. Conversely, optimizing only for the score function on the right side of the plot improves the coverage, albeit with a decrease in the score values.

\subsection{Concentration Threshold Benchmark}

\begin{figure}[!ht]
    \centering
	\includegraphics[width=\linewidth]{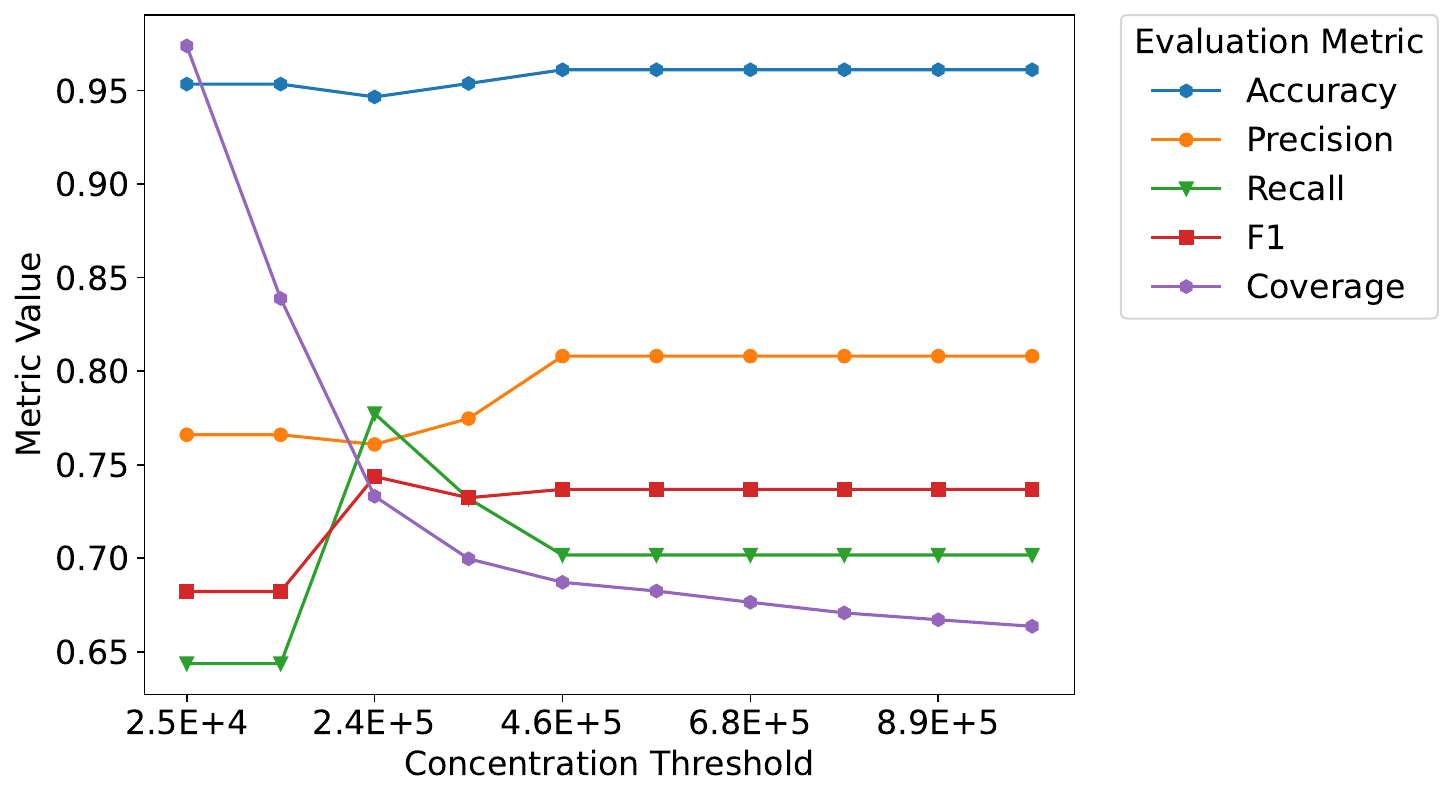}
	\caption{Concentration Threshold Benchmark.}
\label{fig:conc_thresh_benchmark}
\end{figure}

We also benchmarked our solution quality with different concentration values in the threshold indicator function (Equation~\ref{ind_fn_conc}). We used the naive greedy optimizer~\cite{Minoux78} and generated solutions for $6$ sensor placements. Figure~\ref{fig:conc_thresh_benchmark} shows our results. 

As we had anticipated, using low concentration thresholds enables us to detect a greater portion of the outbreak scenarios. This is due to the fact that the sensor placements are closer to the root node, where all wastewater ultimately flows. Consequently, detecting most virus outbreaks is feasible as the low virus concentration is not an issue. However, as we increase the concentration threshold, sensors must be repositioned closer to the leaf nodes (i.e., buildings) to ensure that the wastewater samples contain high virus concentrations. As a result, we would require more sensors to cover all the buildings. Nevertheless, this has the added benefit of improved source localization as each sensor is allocated to a smaller subgraph with fewer leaf nodes.

\begin{figure*}[!ht]
    \centering
	\includegraphics[width=0.8\linewidth]{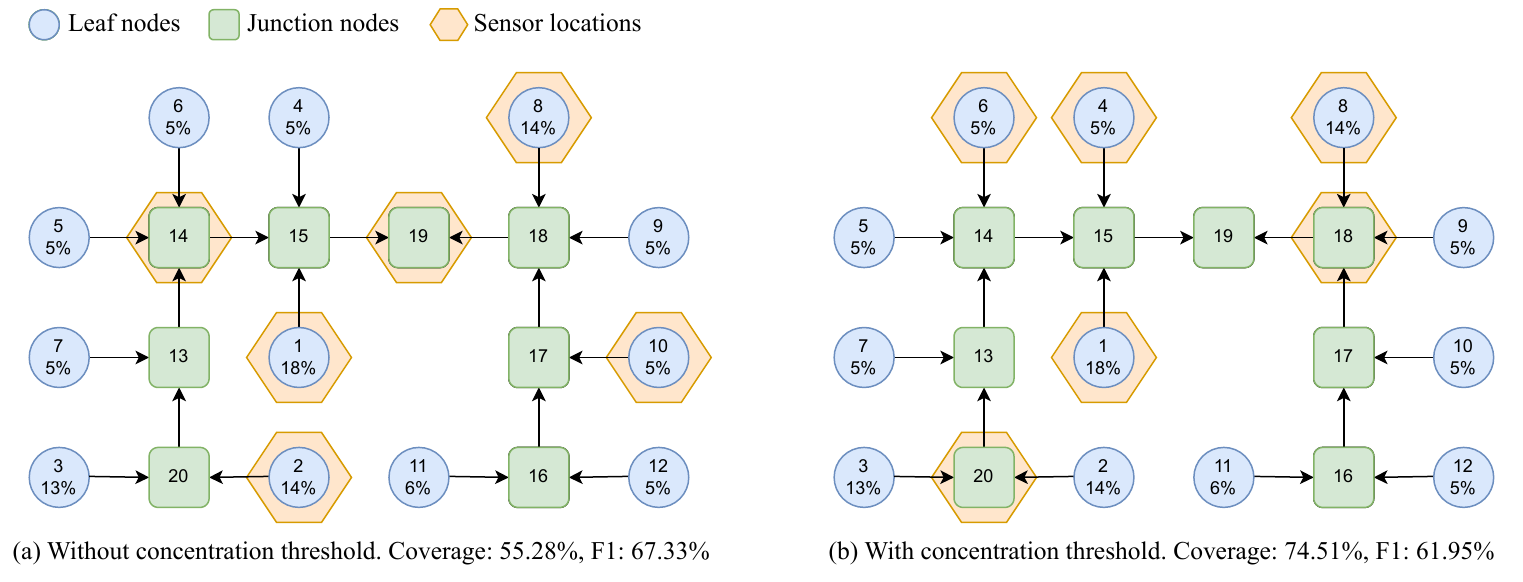}
	\caption{Solution placements for a subgraph of our university's wastewater network. (a) Solution computed without a concentration threshold. (b) Solution computed with a concentration threshold of $4.8 \times 10^5$ virus copies per liter. The percentages in the leaf nodes (blue circles) indicate the fraction of the total network population at each node. The solution placements were evaluated with a concentration threshold of $4.8\times10^{5}$ virus copies per liter.}
 
\label{fig:sovi}
\end{figure*}

Figure~\ref{fig:sovi} illustrates the solution placements on our university's wastewater network graph. We observe that the solution generated without a concentration threshold (Figure~\ref{fig:sovi}(a)) provides inadequate coverage (55.28\%) as the wastewater at the root node (i.e., node 19) is too diluted to produce reliable virus detection results. However, the solution generated with a threshold of $4.8\times10^{5}$ virus copies per liter (Figure~\ref{fig:sovi}(b)) delivers substantially improved coverage (74.51\%).

\subsection{Detection Threshold Benchmark}

\begin{figure}[!ht]
    \centering
	\includegraphics[width=\linewidth]{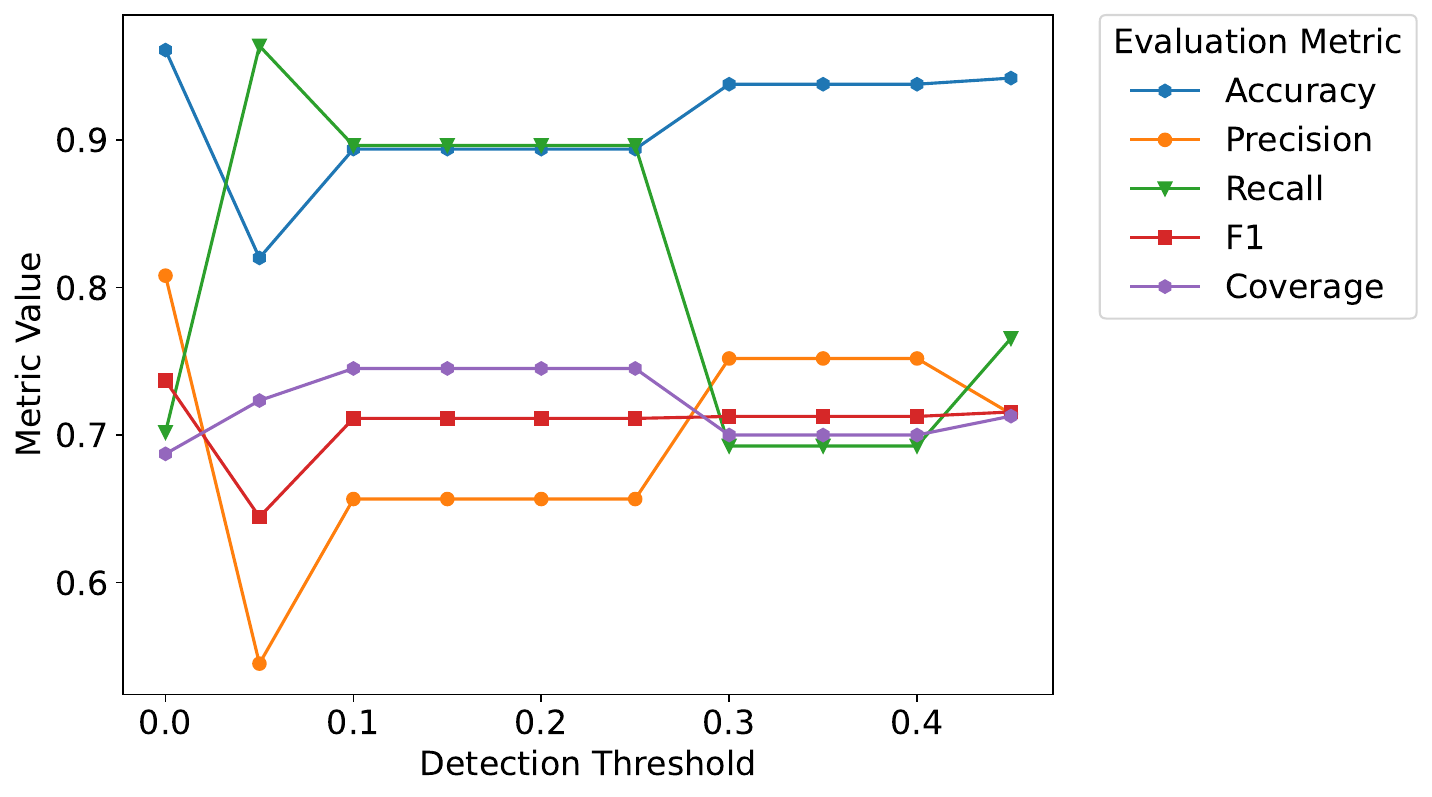}
	\caption{Probability Detection Threshold Benchmark.}
\label{fig:detection_thresh_benchmark}
\end{figure}

In another experiment, we evaluated the impact of varying the virus detection threshold. To determine the virus outbreak state of each building, we thresholded the state probabilities (Equation~\ref{preds}), i.e., label the state as $True$ if the probability is above the threshold, even if the $False$ probability is higher than the $True$ probability. Figure~\ref{fig:detection_thresh_benchmark} illustrates the solution quality obtained using different detection thresholds. For all evaluations in this benchmark, we used the same sensor placements acquired in Section~\ref{metric_benchmark} with the F1 score objective. The sensor placements consisted of 6 sensors and were optimized using the naive greedy optimizer~\cite{Minoux78}.

Note that the zero threshold corresponds to using the original argmax operation (Equation~\ref{preds}) to determine the state of each building. The results demonstrate the classic inverse relation between precision and recall. Precision heavily penalizes the score if we predict a building with a $False$ virus outbreak state as $True$  (false positives). On the other hand, recall allows us to make the same prediction without being penalized, but instead, it penalizes the score if we predict a building with a $True$ virus outbreak state as $False$ (false negative). As such, we observe that increasing the detection threshold initially improves recall, as it increases the fraction of true positives at the cost of increased false positives. However, beyond a threshold of 0.25, it improves precision instead, as a larger fraction of predictions become true positives but with increased false negatives.

\subsection{Random Graph Benchmark}

The experiments we conducted previously focused on a subgraph of our university's wastewater network, allowing us to isolate and study the effects of each variable of interest. In this experiment, we aimed to test the generalizability of our approach to different wastewater networks. To do so, we generated 20 random wastewater network graphs and their corresponding outbreak scenarios, and then generated sensor placement solutions for each graph.

Our random graphs consisted of 25 nodes each. The graphs were built by iteratively linking a new node to a randomly selected existing node. Additionally, we redirected new nodes to successor nodes of the randomly selected nodes, away from the root with a probability of $20\%$. We then sampled the building populations from a uniform distribution in the interval $[0, 100]$, and the wastewater flows for each leaf node from a uniform distribution in the interval $[1000, 3000]$.

For each graph, we sampled 500 virus outbreak scenarios and optimized the placement of 6 sensors using the naive greedy algorithm. The thresholded coverage objective with the F1 score (Equation~\ref{opt_ind_fn_conc}) was used as the optimization objective. We set the concentration threshold to $4.8\times10^{5}$ virus copies per liter.

\begin{figure}[!ht]
    \centering
	\includegraphics[width=\linewidth]{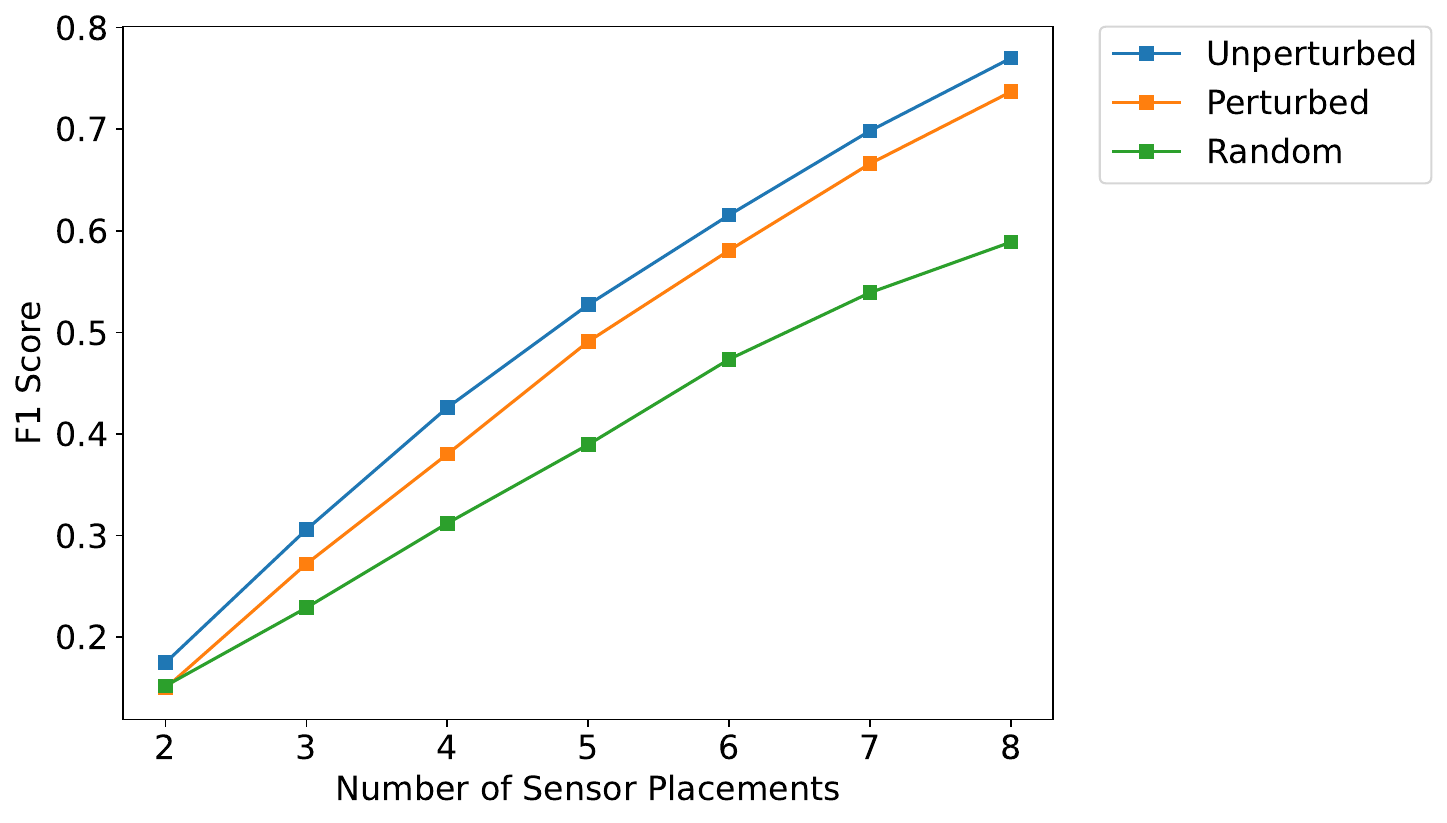}
	\caption{F1 scores on the Random Graph Benchmark.}
\label{fig:random_graph_benchmark_f1}
\end{figure}

\begin{figure}[!ht]
    \centering
	\includegraphics[width=\linewidth]{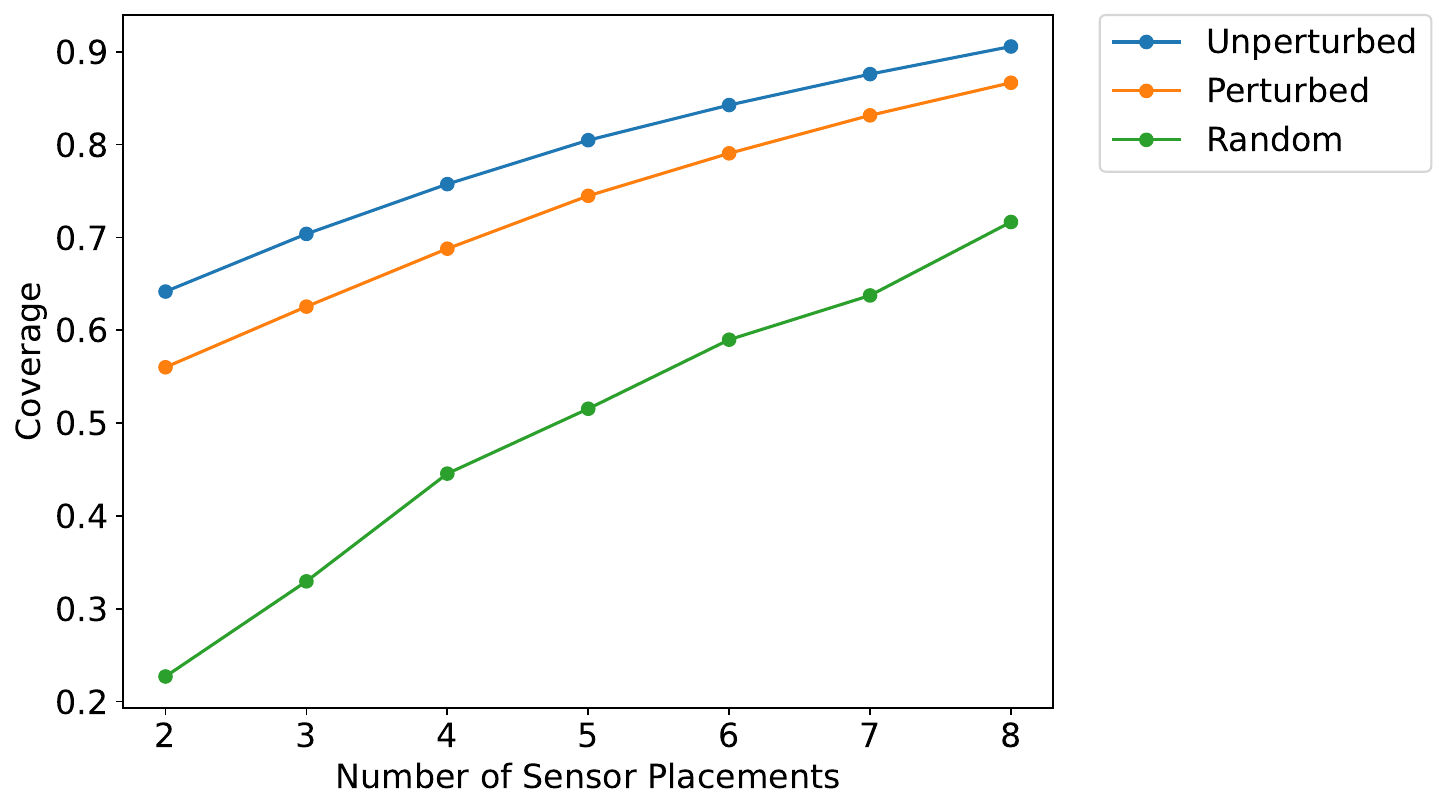}
	\caption{Coverage rates on the Random Graph Benchmark.}
\label{fig:random_graph_benchmark_coverage}
\end{figure}

Figure~\ref{fig:random_graph_benchmark_f1} and Figure~\ref{fig:random_graph_benchmark_coverage}~(Unperturbed) show the average solution F1 score and coverage on the 20 random graphs. We compared our results to a baseline of randomly placed sensors (Random) and found that our approach produced substantial improvements in performance, comparable to those obtained on our university's wastewater network. Note that the random graphs have varying connectivity and different ratios of sensor placements to graph size when compared to our university's wastewater network.

To test the limits of our approach further, we perturbed the virus outbreak probability of each building in the random graphs and generated a new set of scenarios for each random graph. We perturbed the virus outbreak probabilities by adding uniform-distributed noise to the populations of each building in the random graphs. We evaluated the previously obtained sensor placement solutions that were optimized on the unperturbed virus outbreak probability-based scenarios using the new scenarios.  This experiment allowed us to test our approach in real-world scenarios where the available virus outbreak probability or the whole model itself may be inconsistent with the actual virus outbreak characteristics. The results, shown in Figure~\ref{fig:random_graph_benchmark_f1} and Figure~\ref{fig:random_graph_benchmark_coverage}~(Perturbed), demonstrate that our approach is robust to modeling inconsistencies.

%% file: conclusion.tex
\section{Conclusion}
We presented the sensor placement for source localization problem in wastewater networks, and developed an approach that leverages graph Bayesian learning and discrete optimization to address the problem.

We first presented an approach to reduce the size of wastewater network graphs, thereby making relatively large problems computationally feasible. We then showed how one can map any network graph to a Bayesian graph, which we can use to localize sources of information, i.e., buildings experiencing a virus outbreak in our case. We also developed optimization objectives that we can use to efficiently find ideal sensor placements for source localization, even when there are multiple pathogen sources and constraints such as wastewater concentration requirements.

Our simulation experiments demonstrated the quality of our solution sensor placements and the accuracy of our source localization approach in a case study on our university's wastewater network. We also benchmarked different discrete optimization methods and score functions, and showed that our optimization objective can be efficiently optimized. We then established the accuracy of our approach on random graphs. In addition, our experiments established the robustness of our approach to inaccurate virus outbreak models.

In a real-world scenario, our graph sensor placement approach coupled with virus outbreak models and information about the wastewater network structure can determine ideal wastewater sampling locations. Our source localization approach can quickly localize the source of virus outbreaks from regularly collected wastewater sample test results. Additionally, our graph reduction and source localization approaches can be used with existing wastewater network sensor placements approaches.

Our experiments have demonstrated that our optimization objective is submodular on our university's wastewater network graph. We conjecture that this property holds for any wastewater network graph, and we plan to prove this theoretically in our future work.

Furthermore, to validate our approach using real-world data, we require regular virus concentration measurements and the number of infected individuals for each building in the monitored wastewater network. We aim to collect such data and present additional validation of our approach in future work.

%% file: appendix.tex
\section{Appendix}
\begin{table}[h]
\centering
\begin{tabular}{p{0.18\linewidth} p{0.7\linewidth}}
    \hline
    Variable & Definition \\
    \hhline{--}
    $G=(V, E)$ & Wastewater network graph $G$, with vertices $V$ and edges $E$\\
    $B$ & OR-gate Bayesian graph \\
    $C$ & Bayesian graph with variables for modeling sample virus concentration \\
    $l \in L$ & A leaf node, corresponds to a building \\
    $j \in J$ & A non-leaf node, corresponds to a junction node in the reduced graph \\
    $W$ & Set of nodes (leaf/non-leaf nodes) where the sensors are placed. \\
    $W_b$ & The virus outbreak state of the sensing nodes $W$\\
    $A_l(W_b)$ & Virus outbreak state of leaf node $l$ given the current virus presence state of sensing nodes $W_b$ \\
    $S$ & Virus outbreak scenario \\ 
    $k$ & Number of sensors available to be deployed \\
    $b$ & Random variable to indicate whether samples from a node contain viruses \\
    $\hat{n}$ & Random variable to model the number of infected individuals in a building \\
    $n$ & Random variable to model the number of total virus copies shed from a building \\
    $f$ & Random variable to model the volume of wastewater flowing from a building \\
    $c$ & Random variable to model the concentration of a wastewater sample collected at a node \\
    \hline
\end{tabular}
\caption{Definitions of variables.}
\label{tab:vars}
\end{table}

\begin{algorithm}[hb]
\label{alg:sparsify}
\caption{Pseudocode for the recursive function to reduce the wastewater network graph by removing redundant nodes using the above detailed approach.}
\SetKw{Break}{Break}
\SetKw{And}{and}
\textbf{Function name:} Sparsify \\
\KwIn{Graph $G$}
\KwOut{Reduced graph $G$}

\For{vertex $v \in V$}{
    \uIf{in-degree$(v)==1$ \And out-degree$(v)==1$}{
        v-parent = parent$(v)$ \;
        v-child = child$(v)$ \;
        Add-edge(v-parent, v-child) \;
        Delete($v$) \;
        $G = \ $sparsify$(G)$ \;
        \Break\;
    }
}
\Return{$G$}
\end{algorithm}